\documentclass[fleqn,twoside]{article}
\usepackage{espcrc2}

\usepackage{graphicx}
\usepackage[figuresright]{rotating}

\newcommand{\AmS}{{\protect\the\textfont2
  A\kern-.1667em\lower.5ex\hbox{M}\kern-.125emS}}

\newcommand{\be}{\begin{equation}}
\newcommand{\ee}{\end{equation}}

\newcommand{\bea}{\begin{eqnarray}}
\newcommand{\eea}{\end{eqnarray}}

\newcommand{\nn}{\nonumber}

\newcommand{\dslash}{D \!\!\!\! /}    % this is the Dslash operator = D_\mu gamm\a_\mu

\title{QCD at non-zero temperature and density from the Lattice}

\author{C.R. Allton\address[swan]
{Department of Physics, University of Wales Swansea, Swansea SA2 8PP, UK.}
\thanks{Presented by C.R. Allton. This work is supported by BMBF grant
No. 06BI102, DFG grant KA 1198/6-4, PPARC grant PPA/G/S/1999/00026 and
KBN grant 2P03 (06925).},
S. Ejiri\address[biel]
{Fakult\"at f\"ur Physik, Universit\"at Bielefeld, D-33615 Bielefeld, Germany.},
S.J. Hands\addressmark[swan],
O. Kaczmarek\addressmark[biel],
F. Karsch\addressmark[biel],
E. Laermann\addressmark[biel] and
Ch. Schmidt\addressmark[biel]
}

\begin{document}

\begin{abstract}
The study of systems as diverse as the cores of neutron stars and heavy-ion
collision experiments requires the understanding of the phase structure
of QCD at non-zero temperature, $T$, and chemical potential, $\mu_q$.
We review some of the difficulties of performing lattice simulations of QCD
with $\mu_q \ne 0$, and outline the re-weighting method used to overcome
this problem.
This method is used to determine the critical endpoint of QCD in the
$(\mu_q,T)$ plane.
We study the pressure and quark number susceptibility at small $\mu_q$.

\vspace{1pc}
\end{abstract}

\maketitle

%}}}

%{{{ Introduction

\section{Introduction}

\begin{figure}[htb]
%\vspace{9pt}
\includegraphics[height=5cm]{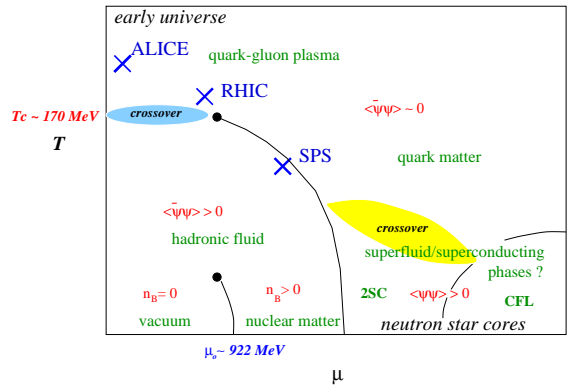}
\caption{The QCD phase diagram in the temperature-chemical potential plane.
First order phase transitions are shown as solid lines, and second
order transitions by filled circles.
Crosses depict heavy-ion collision experiments.
The region of interest in this work is the {\em crossover} transition
at $T_c \approx 170$ MeV.
}
\label{fig:qcd}
\end{figure}

The QCD phase diagram is a rich source of interesting phenomena which
is key to understanding a wide variety of physical systems,
e.g. the early universe, the cores of neutron stars and
heavy-ion collisions experiments.
Figure \ref{fig:qcd} illustrates the QCD phase diagram in the
chemical potential ($\mu_q$) temperature ($T$) plane.
It depicts a very rich structure including familiar phases such as
nuclear matter, together with very unfamiliar ones such as the colour
superconducting phase. See e.g. \cite{satz} for further information.

Lattice simulations of QCD with non-zero chemical potential have
recently provided important qualitative and quantitative information
on the QCD phase diagram in the $(T,\mu_q)$ plane \cite{katz}.
This work will concentrate on studies at finite $T$ and small $\mu_q$
which are directly applicable to heavy-ion collision experiments.
These are shown as crosses in fig \ref{fig:qcd} labelled by the name
of the relevant experiment.

We begin this paper by outlining the difficulties of simulating
QCD with non-zero $\mu_q$, and then define the 
reweighting method which is the basis for much of this work.
The method is then used to determine the crossover
transition near $\mu=0$  between the confining and
deconfining phase (which is depicted as a filled circle
in fig \ref{fig:qcd} near the RHIC point).
We also use this method to determine the pressure and quark
number susceptibility for QCD.
Full details of this work appear in \cite{us} \& \cite{us2}.

%}}}

%{{{ The Sign Problem

\section{The Sign Problem}

In lattice simulations of QCD at $\mu_q=0$, the quark matrix,
${\cal M}(\mu_q=0) = \dslash + m$, has the property
\be
{\cal M}(\mu_q=0) = \gamma_5 {\cal M}^\dagger(\mu_q=0) \gamma_5
\ee
which means that the Boltzmann weight is real, and can be interpreted
as a probability measure.
However, when a non-zero chemical potential is introduced,
the quark matrix becomes (schematically)
${\cal M}(\mu_q) = \dslash + m + \mu_q \gamma_0$. It therefore no longer has
the hermiticity property, and the corresponding Boltzmann weight
is complex (the so-called {\it Sign Problem}),
removing the probabilistic interpretation.
This means that the traditional Monte Carlo approach of estimating
the path integral is a non-starter.

A number of ways have been promoted to overcome this problem.
These fall into two methods. The first is to study {\em models} of QCD
which, due to their simpler mathematical formulation, do not
suffer from the sign problem for $\mu_q \ne 0$.
Such models include the NJL model and 2-colour QCD.
(See \cite{hands} for more details.)

The second approach effectively studies QCD at $\mu_q = 0 + \epsilon$
by using either a reweighting technique
\cite{gottleib,fodorkatz,qcdtaro,us,gavai}
or an analytical continuation approach \cite{defor,lombardo}.
The latter approach simulates with an imaginary $\mu_q$ and
analytically continues back to the real $\mu_q-$axis.
The reweighting technique calculates the expectation value
of some operator, $\left< {\cal O} \right>$,
by factoring off the $\mu_q \ne 0$ part of the Boltzmann weight
and appending it to the observable ${\cal O}$.
Using this approach ensembles generated at $\mu_q=0$ can be used.
Expressing this idea mathematically we have,
\bea \nn
\left<{\cal O}\right>_{(\beta,m,\mu)}
\!\!\!\!\!\!\!\!\!\!\!\!\!\!\!\!\!\!\!\!\!\!\!\!\!\!\!\!\!\!\!\!\!\!
& &
\\ \nn
&=& \frac{1}{{\cal Z}(\beta,m,\mu)}
\int {\cal D}U \;{\cal O}\; det {\cal M}(m,\mu)\; e^{-S_g(\beta)}
\\ \nn
&=& \frac{
 \left<{\cal O} \frac{det{\cal M}(m  ,\mu  ) e^{-S_g(\beta  )}}
                     {det{\cal M}(m_0,\mu_0) e^{-S_g(\beta_0)}}\right>_
{(\beta_0,m_0,\mu_0)}}
{\left<         \frac{det{\cal M}(m  ,\mu  ) e^{-S_g(\beta  )}}
                     {det{\cal M}(m_0,\mu_0) e^{-S_g(\beta_0)}}\right>_
{(\beta_0,m_0,\mu_0)}}
\\ \nn
&=& \frac{
 \left<{\cal O} e^{\ln \left(\frac{det{\cal M}(m  ,\mu  )}
                                  {det{\cal M}(m_0,\mu_0)}\right)
             -S_g(\beta  )
             +S_g(\beta_0)}\right>_{(\beta_0,m_0,\mu_0)}}
{\left<         e^{\ln \left(\frac{det{\cal M}(m  ,\mu  )}
                                  {det{\cal M}(m_0,\mu_0)}\right)
             -S_g(\beta  )
             +S_g(\beta_0)}\right>_{(\beta_0,m_0,\mu_0)}}
\\
&=& \frac{\left<{\cal O} e^{-W}\right>_{(\beta_0,m_0,\mu_0)}}
         {\left<         e^{-W}\right>_{(\beta_0,m_0,\mu_0)}}.
\label{eq:rew}
\eea
In these equations, $S_g$ is the gauge action,
$U$ denotes the gauge degrees of freedom,
$\beta = 6/g^2$, where $g$ is the gauge coupling,
and we have introduced the dimensionless lattice chemical potential,
$\mu = \mu_q a$.
By setting $\mu_0$ to zero, eq(\ref{eq:rew}) enables the calculation
of quantities at $\mu \ne 0$ using ensembles generated at $\mu_0 = 0$.

The results presented in this work are derived from the reweighting
approach as well as a Taylor expansion of $\left<{\cal O}\right>$
about $\mu_q=0$.

%}}}

%{{{ Lattice details

\section{Lattice details}
\label{lattice}

Our simulations were performed on a $N_s^3 \times N_t = 16^3\times 4$
lattice with two (degenerate) quark flavours (see \cite{us} \cite{us2}
for full details).  We used two different quark masses, $ma =$ 0.1 and
0.2 corresponding to pseudoscalar-vector meson mass ratios of
$M_{PS}/M_V \approx$ 0.70 and 0.85 respectively \cite{massratio}.  A
Symanzik improved gauge action together with the {\it p4} improved
quark action \cite{p4} was used to reduce lattice discretisation
artefacts.

A range of $\beta$ values was simulated, each corresponding to a different
temperature via the usual relationship, $T = 1/ (a(\beta) N_t)$, where
$N_t$ is the number of lattice sites in the temporal direction.
${\cal O}(1,000,000)$ trajectories were generated in this project
using the APEmille computers in Bielefeld and Swansea.

%}}}

%{{{ Reweighting preliminaries

\section{Reweighting preliminaries}

\begin{figure}[htb]
%\vspace{9pt}
\includegraphics[height=7cm]{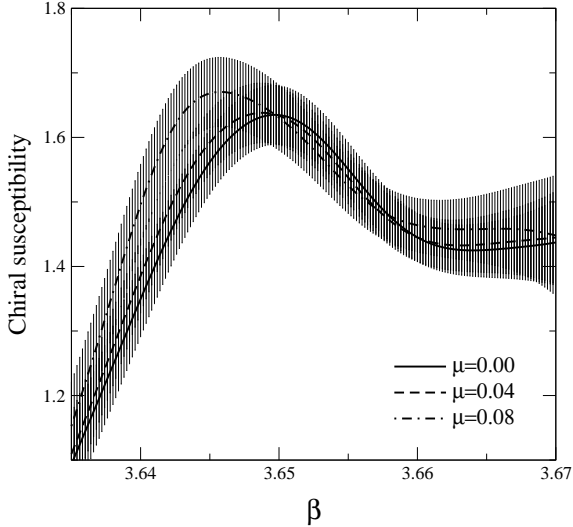}
\caption{The chiral susceptibility, $\chi_{\overline{\psi}\psi}(\beta)$
at $ma = 0.1$ for various $\mu$.
}
\label{fig:rew}
\end{figure}

As an illustration of the reweighting method we show the susceptibility
of the chiral condensate, $\chi_{\overline{\psi}\psi}$ as a function of
$\beta$ in fig \ref{fig:rew} for a quark mass of $ma = 0.1$.
Note that to produce this plot, a reweighting has been performed both
in $\beta$ and $\mu$ (i.e. the simulations were only performed a relatively
coarse values of $\beta$ compared to those shown in fig \ref{fig:rew}).

\begin{figure}[htb]
%\vspace{9pt}
\includegraphics[height=7cm]{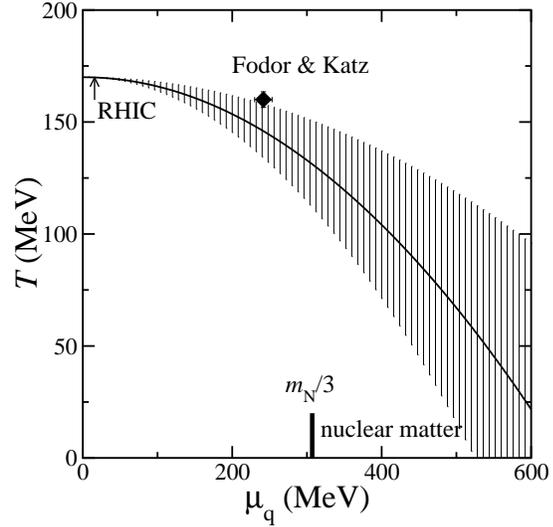}
\caption{The phase diagram obtained from our reweighting technique.
The errors shown are statistical.
The diamond is the endpoint of the first order phase transition obtained
by Fodor and Katz \cite{fodorkatz}.
}
\label{fig:phasediag}
\end{figure}

As can be seen, there is a clear peak in the susceptibility, which
occurs at the crossover transition.
This means that the expectation value of $\overline{\psi} \psi$
fluctuates strongly at this $\beta$ value corresponding
to a temperature $T = 1/(a(\beta)N_t)$.
By mapping out the position of this transition as a function of
$\beta$ and $\mu$, the phase diagram, fig \ref{fig:phasediag},
can be obtained \cite{us}.

Note that figs \ref{fig:rew} and \ref{fig:phasediag} were obtained by
Taylor expanding the susceptibilities to ${\cal O}(\mu^2)$.
The phase diagram, fig \ref{fig:phasediag} is therefore correct
to this order in $\mu$ \cite{us}.

%}}}

%{{{ Pressure

\section{Pressure}

The pressure is related to the grand partition function, ${\cal Z}(T,V,\mu_q)$
by
\be
\frac{p}{T^4} = \frac{1}{VT^3} \ln {\cal Z}.
\label{eq:p1}
\ee
We can now Taylor expand both sides of this equation in the dimensionless
ratio $\mu_q/T$. Since we can only utilize Monte Carlo methods at
$\mu_q\equiv 0$, we perform this expansion about this point obtaining
\bea \nn
\Delta \left( \frac{p}{T^4}(\mu_q) \right) &\equiv&
\frac{p}{T^4}\biggr\vert_{T,\mu_q} - \frac{p}{T^4}\biggr\vert_{T,0} \\ \nn
&\!\!\!\!\!\!\!\!\!\!\!\!\!\!\!\!\!\!\!\!\!\!\!\!\!\!\!\!\!\!\!\!\!\!\!\!\!
     =&\!\!\!\!\!\!\!\!\!\!\!\!\!\!\!\!\!\!\!\!
\frac{1}{2!}\frac{\mu_q^2}{T^2}\frac{\partial^2(p/T^4)}{\partial(\mu_q/T)^2} +
\frac{1}{4!}\frac{\mu_q^4}{T^4}\frac{\partial^4(p/T^4)}{\partial(\mu_q/T)^4} +
\ldots \\
&\!\!\!\!\!\!\!\!\!\!\!\!\!\!\!\!\!\!\!\!\!\!\!\!\!\!\!\!\!\!\!\!\!\!\!\!\!
\equiv&\!\!\!\!\!\!\!\!\!\!\!\!\!\!\!\!\!\!\!\!
 \sum_{n=2,4,\ldots}^{\infty} c_n(T)\;\;\left(\frac{\mu_q}{T}\right)^n,
\label{eq:p2}
\eea
where all derivatives are taken at $\mu_q=0$. Only even powers occur in
eq(\ref{eq:p2}) because odd derivatives of the free energy with respect to
$\mu_q$ vanish \cite{us}.
Using eq(\ref{eq:p1}), the derivatives in eq(\ref{eq:p2}) can be expressed as
expectation values (over a $\mu_q=0$ ensemble).

\begin{figure}[htb]
%\vspace{9pt}
\includegraphics[height=7cm]{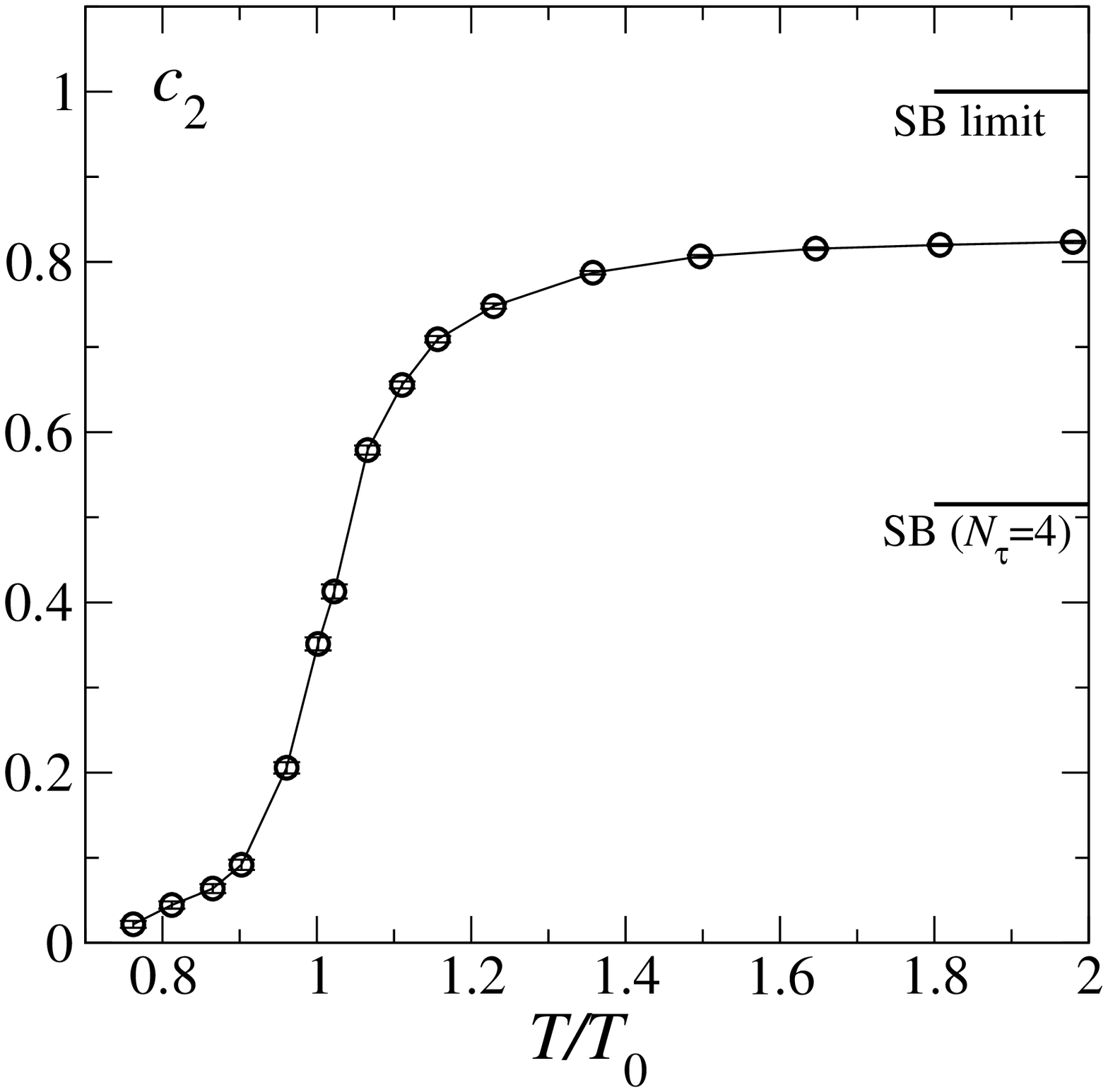}
\caption{The quadratic coefficient of the Taylor expansion of the pressure,
$c_2$, as a function of temperature. (See eq(\ref{eq:p2}).)
}
\label{fig:c2}
\end{figure}

\begin{figure}[htb]
%\vspace{9pt}
\includegraphics[height=7cm]{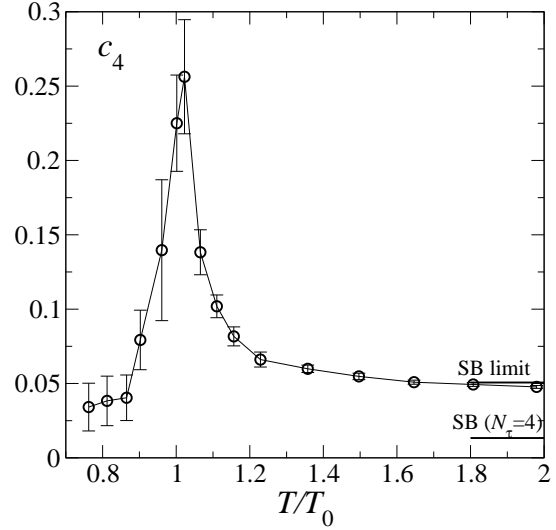}
\caption{The quartic coefficient of the Taylor expansion of the pressure,
$c_4$, as a function of temperature. (See eq(\ref{eq:p2}).)
}
\label{fig:c4}
\end{figure}

In figs \ref{fig:c2} and \ref{fig:c4} the first two coefficients of $p$,
$c_{2,4}$, are plotted as a function of temperature.
For comparison, the Stefan-Boltzmann (SB) values are plotted.
These are valid in the limit of infinite temperature, where the system
becomes asymptotically free.
The continuum SB expression for massless quarks is
\be \nn
\frac{p_{SB}}{T^4}(\mu_q) = \frac{7N_f \pi^2}{60}
 + \frac{N_f}{2}\left(\frac{\mu_q}{T}\right)^2
 + \frac{N_f}{4\pi^2}\left(\frac{\mu_q}{T}\right)^4,
\ee
so the coefficients $c_{2,4}$ can be simply read-off.
Also shown in figs \ref{fig:c2} and \ref{fig:c4} are SB values
corresponding to a free field lattice calculation at $N_t = 4$
(i.e. corresponding to the temporal extent of our lattice,
see sec. \ref{lattice}).
The details of this calculation can be found in \cite{us2}.
As can be seen from figs \ref{fig:c2} and \ref{fig:c4}, both $c_2$ and
$c_4$ vary sharply in the critical region, $T\approx T_0$.

\begin{figure}[htb]
%\vspace{9pt}
\includegraphics[height=7cm]{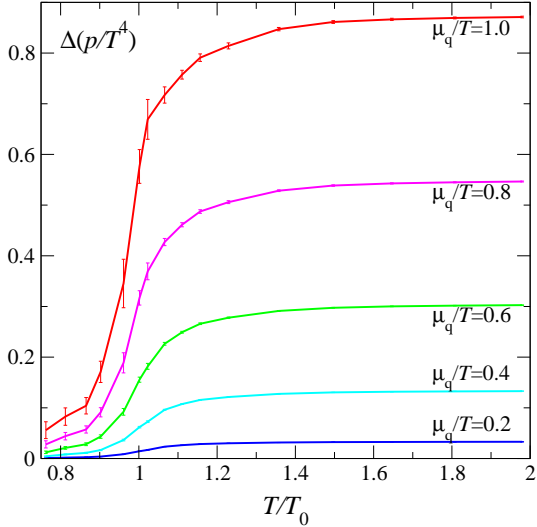}
\vspace{-11pt}
\caption{The dimensionless pressure difference $\Delta(p/T^4)$
as a function of temperature for various $\mu_q/T$ values.
}
\label{fig:deltap}
\end{figure}

Using the coefficients $c_{2,4}$, we plot the pressure difference,
$\Delta (p/T^4)$ as a function of temperature for various $\mu_q/T$ in
fig \ref{fig:deltap}.
Again we observe an abrupt change in the pressure as the critical
region is crossed.
It is interesting to note that the pressure 
at finite density is very similar to its $\mu_q=0$ value.
For instance, at the RHIC point ($\mu_q/T \approx 0.1$) the pressure
is only 1\% more than its zero density value.

%}}}

%{{{ Susceptibility

\section{Susceptibility}

\begin{figure}[htb]
%\vspace{9pt}
\includegraphics[height=7cm]{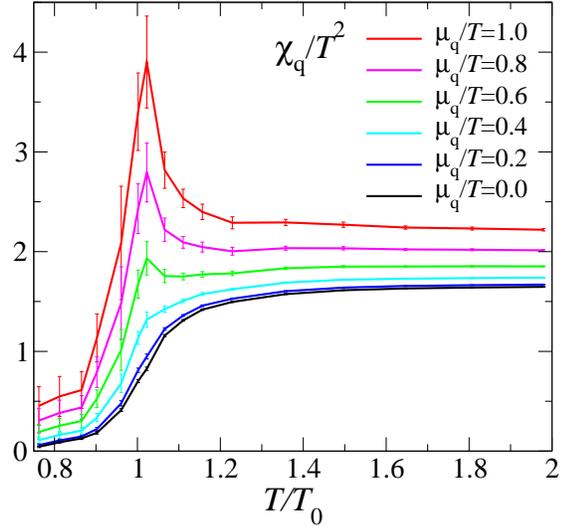}
\vspace{-11pt}
\caption{The quark number susceptibility, $\chi_q$, as a function of
temperature, for various $\mu_q/T$ values.
}
\label{fig:sus}
\end{figure}

We turn our discussion to the quark number density, $n_q$, and its
susceptibility, $\chi_q$, defined as,
\bea \nn
\frac{n_q}{T^3}    &=& \frac{\partial  (p/T^4)}{\partial(\mu_q/T)} \\ \nn
\frac{\chi_q}{T^2} &=& \frac{\partial^2(p/T^4)}{\partial(\mu_q/T)^2}. \\ \nn
\eea
The ingredients required to calculate the susceptibility have already
been determined in the Taylor expansion of the pressure
(see eqs(\ref{eq:p1} \& \ref{eq:p2})).
Figure \ref{fig:sus} plots $\chi_q/T^2$ as a function of temperature
for various $\mu_q/T$ values.
As can be seen, a peak in $\chi_q$ develops at $T \approx T_0$ which
grows with $\mu_q$.
This is a signal of a critical endpoint in the $(\mu_q,T)$ plane
where the fluctuations in the number operator $\overline{\psi} \gamma_0 \psi$
grow.
Also note that the peak occurs at $T=T_0$, implying that the chiral symmetry
restoration transition observed in fig \ref{fig:rew} and the deconfinement
transition of fig \ref{fig:sus} coincide \cite{satz}.

%}}}

%{{{ Conclusions

\section{Conclusions}

We have presented some results of studies of QCD at non-zero temperature
and small chemical potential \cite{us,us2}.
In particular we have summarised the difficulties in using conventional
Monte Carlo techniques to perform lattice simulations of QCD at non-zero
$\mu$.
We outlined the reweighting technique which has recently been very
successful in obtaining numerical results at $\mu\ne 0$, despite
this difficulty \cite{fodorkatz,us}.
This method was used to determine the position of the crossover transition
between the confined and de-confined phases.
Finally, results for the pressure and susceptibility were produced, based on a
Taylor expansion in $\mu_q/T$.

%}}}

%{{{ bibliography

%}}}

\end{document}